\documentclass[a4paper,10pt]{article}

\title{Langevin Approach to Fractional Diffusion Equations including Inertial Effects}
\author{R.~Friedrich, S.~Eule\footnote{yeahhart@uni-muenster.de, Tel.:+49-251-8334924, Fax: +49-251-8336328}, F.~Jenko}

\begin{document}

\maketitle
\centerline{Institute of Theoretical Physics}
\centerline{Westf\"alische Wilhelms Universit\"at}
\centerline{M\"unster}
\centerline{Wilhelm-Klemm-Str. 9}
\centerline{G-48149 M\"unster}

\begin{abstract}
In recent years, several fractional generalizations of the usual 
Kramers-Fokker-Planck equation
have been presented.
Using an idea of Fogedby [H.C. Fogedby, Phys. Rev. E {\bf 50}, 041103 (1994), 
we show how these equations are related to 
Langevin equations via the procedure of subordination.
\end{abstract}

{\em Introduction.} -- Some 70 years ago, Kramers \cite{Kramers} considered the motion of a
Brownian particle subject to a space-dependent force ${\bf F}({\bf x})$ per unit mass. His
goal was to compute the joint probability distribution $f({\bf x},{\bf u},t)$ for finding a
particle at time $t$ at the position ${\bf x}$ with the velocity ${\bf u}$. For this quantity
he could derive the famous Kramers-Fokker-Planck (KFP) equation \cite{Risken,vanKampen}
\begin{equation}\label{Kramers}
   \left[{\partial\over\partial t}+{\bf u}\cdot\nabla_x + {\bf F}({\bf x})\cdot\nabla_u \right]
   f({\bf x},{\bf u},t) = {\cal L}_{\rm FP}f({\bf x},{\bf u},t)
\end{equation}
where ${\cal L}_{\rm FP}$ is the Fokker-Planck collision operator
\begin{equation}\label{FPop}
   {\cal L}_{\rm FP}f=\gamma\nabla_u\cdot({\bf u}f)+D\,\Delta_u f\,.
\end{equation}
As is well known, Eq.~(\ref{Kramers}) corresponds to the Langevin equations (see, e.g.,
Ref.~\cite{Risken})
\begin{equation}\label{dyn1}
  {d\over dt}\,{\bf x}(t) = {\bf u}(t)\,,\quad
  {d\over dt}\,{\bf u}(t) = {\bf F}({\bf x)}-\gamma{\bf u}(t)+\mbox{\boldmath$\Gamma$(t)}
\end{equation}
where $\mbox{\boldmath$\Gamma$(t)}$ obeys white noise statistics. It describes a Brownian
particle which is subject to the relation $\langle {\bf x}^2\rangle\sim D\,t$ where $D$ is the
diffusion coefficient. In many complex systems, this relation is violated, however. In fact,
one often finds $\langle {\bf x}^2\rangle \sim D_\alpha t^\alpha$
with $\alpha\ne 1$ which is described as "anomalous" or "strange" diffusion. Here, $D_\alpha$
is a generalized diffusion coefficient with units $[D_\alpha]=m^2 s^{-\alpha}$. Depending on
$\alpha$, such a process is called subdiffusive ($\alpha<1$), superdiffusive ($1<\alpha<2$),
ballistic ($\alpha=2$), or turbulent-diffusive ($\alpha=3$).

As was shown by several authors, strange diffusion may be described by fractional generalizations
of Eq.~(\ref{Kramers}) (for a review including discussions of various applications see e.g., Ref.~\cite{Coffey}). 
However, the latter may differ in the way the fractional character is incorporated. Thus it comes as no surprise that three different
types of fractional KFP equations may be found in the literature. E.g., Metzler and Klafter
\cite{Metzler3,Metzler3a,Metzler4} proposed the equation
\begin{eqnarray}\label{FFPmk}
   {\partial f({\bf x},{\bf u},t)\over\partial t} = 
   \left[ -{\bf u}\cdot\nabla_x - {\bf F}({\bf x})\cdot\nabla_u + {\cal L}_{\rm FP} \right]
   \gamma_\delta D_t^{1-\delta} f({\bf x},{\bf u},t)
\end{eqnarray}
which they obtained by means of a non-Markovian generalization of the Chapman-Kolmogorov equation.
Another kind of fractional KFP equation has been proposed by Barkai and Silbey \cite{BaSil},
namely
\begin{equation}\label{BSeq}
   \left[{\partial\over\partial t}+{\bf u}\cdot\nabla_x + {\bf F}({\bf x})\cdot\nabla_u \right]
   f({\bf x},{\bf u},t) = {\cal L}_{\rm FP} \gamma_\delta D_t^{1-\delta} f({\bf x},{\bf u},t)
\end{equation}
where $\gamma_\delta$ is a damping coefficient whose units are $[\gamma_\delta]=s^{\delta-1}$
and $D_t^{1-\delta}$ is the fractional time derivative whose Laplace space representation
reads $D_t^{1-\delta}\leftrightarrow\lambda^{1-\delta}$. Finally, employing the concept of
continuous time random walks (CTRWs), Friedrich and co-workers \cite{Eule,CTRW} were able
to derive the equation
\begin{eqnarray}\label{Kramers2}
   \left[{\partial\over\partial t}+{\bf u}\cdot\nabla_x
   +{\bf F}({\bf x})\cdot\nabla_u\right] f({\bf x},{\bf u},t) =
   {\cal L}_{\rm FP} \gamma_\delta {\cal D}_t^{1-\delta} f({\bf x},{\bf u},t)\,.
\end{eqnarray}
Here, ${\cal D}_t^{1-\delta}$ denotes a fractional {\em substantial} derivative which can be
written in Laplace space as ${\cal D}_t^{1-\delta} \leftrightarrow [\lambda+{\bf u}\cdot
\nabla_x+{\bf F}({\bf x})\cdot\nabla_u]^{1-\delta}$.

In this letter we address the important question of how these three fractional KFP equations
are connected to each other. As is well known, fractional diffusion equations can be linked
to CTRWs.\cite{Metzler} And according to Fogedby \cite{Fogedby}, the latter can in turn be
linked to sets of Langevin equations. This fact will be exploited below in order to gain
insight into the nature of the stochastic processes underlying the three scenarios that were
just described. For the sake of simplicity, we restrict ourselves to one-dimensional problems
with no external force.

{\em Langevin approach to fractional diffusion equations.} -- In the spirit of Fogedby, let us first
consider a stochastic process which is described by the following system of Langevin equations:
\begin{eqnarray}
   \frac{d}{ds}\,u(s) = -\gamma u(s) + \Gamma(s)\,,\quad\frac{d}{ds}\,t(s) = \eta(s)\,.
\end{eqnarray}
Here, the variable $s$ is to be interpreted as an internal time, whereas $t$ is the physical
(wall-clock) time. Moreover, $\Gamma$ and $\eta$ are described, respectively, by Gaussian and
one-sided L\'evy stable distributions [denoted by $L_\delta(x)$] \cite{Levy}. Mathematically
speaking, the stochastic process $u(s)$ is subordinated by the $t(s)$ process. The latter is
invertible, and the probability density of finding the internal time $s$ at time $t$ is given by
\begin{equation}
   p(s,t) \propto {d\over ds}\,\left[ 1-L_\delta\left(t/(\gamma_\delta s)^{1/\delta}\right) \right]
\end{equation}
which is called the inverse one-sided L\'evy stable distribution. It is a solution of the equation
\begin{equation}\label{Levpdf}
   \frac{\partial}{\partial t}\,p(s,t)=-\frac{\partial}{\partial s}\,\gamma_\delta
   D_t^{1-\delta} p(s,t)\,,
\end{equation}
and its Laplace transform reads $\hat p(s,\lambda)\propto\lambda^{\delta-1}\exp(-\gamma_\delta
\lambda^\delta s)$. Assuming that the
stochastic processes $t(s)$ and $u(s)$ are statistically independent, the probability $P(u,t)$
of finding the velocity $u$ at time $t$ can be written as
\begin{equation}\label{eq1}
   P(u,t) = \int_0^\infty P_0(u,s)\,p(s,t)\,ds
\end{equation}
where the distribution function $P_0(u,s)$ is a solution of the standard diffusion equation,
\begin{eqnarray}\label{eq2}
   {\partial P_0(u,s)\over\partial s} = {\cal L}_{\rm FP} P_0(u,s)\,.
\end{eqnarray}
From Eqs.~(\ref{Levpdf})-(\ref{eq2}) it then follows that $P(u,t)$ satisfies the fractional
diffusion equation
\begin{eqnarray}\label{eq3}
   {\partial P(u,t)\over\partial t} =
   {\cal L}_{\rm FP} \gamma_\delta D_t^{1-\delta} P(u,t)\,.
\end{eqnarray}
The idea of representing the solution of a fractional diffusion equation like Eq.~(\ref{eq3})
as a superposition of Gaussians goes back to Barkai \cite{Barkai}. In the following, we will
extend this method from velocity space to phase (position-velocity) space.

{\em The fractional KFP equation by Metzler and Klafter.} -- Let us now consider the Langevin system
\begin{eqnarray}\label{hilfersys}
   \frac{d}{ds}\,x(s) = u(s)\,,\quad\frac{d}{ds}\,u(s)=-\gamma u(s)+\Gamma(s)\,,\quad
   \frac{d}{ds}\,t(s) = \eta(s)
\end{eqnarray}
which is closely related to Eq.~(\ref{dyn1}). Here, both $x(s)$ and $u(s)$ are subordinated by the
same $t(s)$ process. In analogy with Eq.~(\ref{eq1}), the probability distribution $f(x,u,t)$ can
be written as
\begin{equation}
   f(x,u,t) = \int_0^\infty f_0(x,u,s)\,p(s,t)\,ds
\end{equation}
where $f_0(x,u,s)$ is the solution of the Fokker-Planck equation
\begin{equation}\label{markovfpeq}
   \left[ \frac{\partial}{\partial s} +u \frac{\partial}{\partial x} \right] f_0(x,u,s) =
   {\cal L}_{\rm FP} f_0(x,u,s)\,.
\end{equation}
Using these relations together with Eq.~(\ref{Levpdf}), one obtains
\begin{equation}
   \frac{\partial f(x,u,t)}{\partial t} =
   \left[ - u\frac{\partial}{\partial x} + {\cal L}_{\rm FP} \right]
   \gamma_\delta D_t^{1-\delta} f(x,u,t)\,.
\end{equation}
This is the fractional generalization of the usual KFP equation considered by Metzler and Klafter
\cite{Metzler3,Metzler3a,Metzler4}. Thus we have shown that the corresponding stochastic process
is given by Eq.~(\ref{hilfersys}).

{\em The fractional KFP equation by Barkai and Silbey.} -- Next, we want to consider the Langevin system
\begin{eqnarray}\label{langevinbarkai}
   \frac{d}{dt}\,x(t) = u(t)\,,\quad\frac{d}{ds}\,u(s)=-\gamma u(s)+\Gamma(s)\,,\quad
   \frac{d}{ds}\,t(s) = \eta(s)\,.
\end{eqnarray}
Here, the velocity coordinate is subordinated by the internal time process while the evolution
of the space variable is in physical time. These equations may be rewritten as
\begin{eqnarray}\label{eq5}
    \frac{d}{ds}\,x(s) = u(s)\,\eta(s)\,,\quad\frac{d}{ds}\,u(s)=-\gamma u(s)+\Gamma(s)
    \,,\quad\frac{d}{ds}\,t(s) = \eta(s)\,.
\end{eqnarray}
For any specific realization of $\eta(s)$, one can view this as a stochastic process which
only depends on the Gaussian variable $\Gamma(s)$. The corresponding probability distribution
$f_0(x,u,s)$ is subject to the KFP-type equation
\begin{equation}
   \left[ \frac{\partial }{\partial s}+ u\eta(s)\frac{\partial}{\partial x}
   \right] f_0(x,u,s)={\cal L}_{\rm FP} f_0(x,u,s)\,.
\end{equation}
The solution of this equation is a Gaussian probability distribution with the second order
moments defined by 
\begin{eqnarray}
   \frac{d}{ds} \langle u^2\rangle (s) &=& -2\gamma\,\langle u^2\rangle (s)+2D\,,\nonumber \\
   \frac{d}{ds} \langle xu\rangle (s) &=& -\gamma\,\langle xu\rangle (s) +
   \eta(s)\,\langle u^2\rangle (s)\,, \nonumber \\
   \frac{d}{ds} \langle x^2\rangle (s) &=& 2\eta(s)\,\langle xu\rangle (s)\,.
\end{eqnarray}

For simplicity, we first consider the case $\gamma=0$ in which one obtains
\begin{eqnarray}
   \langle u^2\rangle (s(t)) &=& 2D\,s(t)\,,\nonumber \\
   \frac{d}{dt} \langle xu\rangle (s(t)) &=& \langle u^2\rangle (s(t))\,, \nonumber \\
   \frac{d}{dt} \langle x^2\rangle (s(t)) &=& 2\,\langle xu\rangle (s(t))\,.
\end{eqnarray}
Introducing the auxiliary variables $\sigma(t)$ and $\Sigma (t)$ via
\begin{eqnarray}\label{internal}
   \frac{d}{dt}\,\sigma(t) = s(t)\,,\quad\frac{d}{dt}\,\Sigma(t) = \sigma(t)\,,
\end{eqnarray}
one finds
\begin{eqnarray}
   \langle u^2\rangle (s(t)) = 2D\,s(t)\,,\quad
   \langle xu\rangle (s(t)) = 2D\,\sigma(t)\,,\quad
   \langle x^2\rangle (s(t)) = 4D\,\Sigma(t)\,,
\end{eqnarray}
and the characteristic function 
\begin{equation}
   Z(k,\alpha,\cdot)=\int dx \int du \, f(x,u,\cdot)\,\exp[ikx+i\alpha u]
\end{equation}
of $f_0$ is obtained as 
\begin{equation}\label{charac}
   Z_0(k,\alpha,s,\sigma,\Sigma) = \exp\left[-D\alpha^2\,s(t)
   -2D \alpha k \, \sigma(t)-2D k^2 \, \Sigma(t)\right]\,.
\end{equation}
[We note that the latter can also be calculated directly from
Eq.~(\ref{langevinbarkai}).]

Now, the stochastic process $\eta(s)$ defines a probability distribution
$W(s,\sigma,\Sigma,t)$ where $s(t)$, $\sigma(t)$, and $\Sigma(t)$ are related
to $\eta(s)$ via Eqs.~(\ref{eq5}) and (\ref{internal}). We assume that this
function satisfies the equation
\begin{equation}
   \left[\frac{\partial }{\partial t}+\sigma\frac{\partial}{\partial\Sigma}
   +s\frac{\partial}{\partial\sigma}\right] W(s,\sigma,\Sigma,t) =
   -\frac{\partial }{\partial s} \gamma_\delta D_t^{1-\delta} W(s,\sigma,\Sigma,t)
\end{equation}
which is the natural generalization of Eq.~(\ref{Levpdf}). In analogy with Eq.~(\ref{eq1}),
the generic characteristic function of $f(x,u,t)$ can thus be written as
\begin{equation}\label{zansatz}
   Z(k,\alpha,t)=\int ds \int d\sigma \int d\Sigma \,\,Z_0(k,\alpha,s,\sigma,\Sigma)\,
   W(s,\sigma,\Sigma,t)\,.
\end{equation}
It is straightforward to show that the corresponding distribution function $f(x,u,t)$
obeys the fractional KFP equation
\begin{equation}\label{eq6}
   \left[ \frac{\partial}{\partial t} + u\frac{\partial}{\partial x} \right] f(x,u,t) =
    {\cal L}_{\rm FP} \gamma_\delta D_t^{1-\delta} f(x,u,t)
\end{equation}
with $\gamma=0$.

The case $\gamma \ne 0$ is only slightly more difficult. Eq.~(\ref{eq6}) yields
\begin{equation}
   \left[ \frac{\partial }{\partial t}
   - k \frac{\partial }{\partial \alpha} \right] Z(k,\alpha,t)
   = - \left[ \gamma \alpha \frac{\partial }{\partial \alpha }
   +D\alpha^2 \right] \gamma_\delta D_t^{1-\delta} Z(k,\alpha,t)\,.
\end{equation}
Using Eq.~(\ref{charac}), we note that
\begin{equation}\label{abl1}
   \left[k\frac{\partial}{\partial\alpha}\right]Z_0=\left[\sigma\frac{\partial}{\partial\Sigma}+
   s\frac{\partial}{\partial\sigma} \right]Z_0
\end{equation}
and
\begin{equation}\label{abl2}
\left[\gamma \alpha \frac{\partial}{\partial \alpha} + D \alpha^2 \right] Z_0=
\left[(2\gamma s-1)\frac{\partial}{\partial s}+\gamma\sigma\frac{\partial}{\partial\sigma}\right]Z_0\,.
\end{equation}
By means of Eq.~(\ref{zansatz}), we then obtain
\begin{equation}\label{eq8}
   \left[ \frac{\partial}{\partial t}+\sigma\frac{\partial}{\partial\Sigma}
     +s\frac{\partial}{\partial\sigma} \right] W(s,\sigma,\Sigma,t) = 
    \left[ \frac{\partial}{\partial s}(2\gamma s-1)
    +\gamma \frac{\partial}{\partial\sigma} \sigma \right] \gamma_\delta D_t^{1-\delta}
    W(s,\sigma,\Sigma,t)
\end{equation}
which is a generalization of Eq.~(\ref{Levpdf}). So, starting with Eq.~(\ref{langevinbarkai})
and assuming that the stochastic process $\eta(s)$ defines a probability distribution
$W(s,\sigma,\Sigma,t)$ which satisfies Eq.~(\ref{eq8}), the fractional KFP equation \`a la
Barkai and Silbey [Eq.~(\ref{eq6})] holds.

{\em The fractional KFP equation by Friedrich et al.} -- Finally we address the
generalized KFP equation with retardation proposed in Ref.~\cite{Eule}. It reads
\begin{equation}\label{eq9}
   \left[ \frac{\partial}{\partial t} + u\frac{\partial}{\partial x} \right] f(x,u,t) =
    {\cal L}_{\rm FP} \gamma_\delta {\cal D}_t^{1-\delta} f(x,u,t)
\end{equation}
where ${\cal D}_t^{1-\delta}$ is the fractional {\em substantial} derivative introduced above.
According to Ref.~\cite{CTRW}, this equation can also be written as
\begin{equation}
   \left[\frac{\partial}{\partial t}+u\frac{\partial}{\partial x}\right] f(x, u, t) =
   {\cal L}_{\rm FP} \int_0^t Q(t-t')\, e^{-(t-t')u\partial x}\,f(x,u,t')\,dt'
\end{equation}
if the memory kernel $Q(t-t')$ is chosen appropiately. In this case, the characteristic function
satisfies the equation
\begin{eqnarray}
   & & \left[\frac{\partial}{\partial t}-k\frac{\partial}{\partial\alpha}\right]
   Z(k,\alpha,t) = \nonumber \\
   & & -\int_0^t Q(t-t')\,\left[\gamma\alpha\frac{\partial}{\partial\alpha} + D\alpha^2 \right]
   Z(k,\alpha+k(t-t'),t')\,dt'\,,
\end{eqnarray}
and one finds
\begin{eqnarray}
   Z_0(k,\alpha+k(t-t'),s,\sigma,\Sigma) =
   \exp\left[-D\alpha^2s-2D\alpha k\tilde\sigma - 2 D k^2 \tilde\Sigma \right]
\end{eqnarray}
where we have introduced the new variables ${\tilde\sigma}=\sigma+s(t-t')$ and
${\tilde\Sigma}=\Sigma+\sigma(t-t')+s(t-t')^2/2$. Using relations similar to those in
Eqs.~(\ref{abl1}) and (\ref{abl2}) as well as the ansatz (\ref{zansatz}), we obtain
the evolution equation
\begin{eqnarray}\label{eq11}
\lefteqn{
   \left[ \frac{\partial}{\partial t} + \sigma\frac{\partial}{\partial\Sigma}
   + s\frac{\partial}{\partial\sigma}\right] W(s,\sigma,\Sigma,t)  
}\nonumber \\
&&=
   \int_0^t Q(t-t')\,\left[\frac{\partial}{\partial s}(2\gamma s-1)
   +\gamma\frac{\partial}{\partial\sigma}\sigma\right] \times
\nonumber \\
&&\times
 W(s,\sigma-s(t-t'),\Sigma-\sigma(t-t')+
   \frac s2(t-t')^2,t')\,dt'
\end{eqnarray}
for $W(s,\sigma,\Sigma,t)$. The latter is simply a retarded version of Eq.~(\ref{eq8}).
To clarify the difference between Eq.~(\ref{eq8}) and 
Eq.~(\ref{eq11}) we introduce the shifted variables $\hat{\sigma}=\sigma -st$ and $\hat{\Sigma}=\Sigma -\sigma t+
\frac s2 t^2$  in the sense that
\begin{equation}
W(s,\sigma, \Sigma, t)=\tilde{W}(s, \sigma -st, \Sigma-\sigma t+\frac{s}{2}t^2,t)
\end{equation} 
holds. Consequently Eq. (\ref{eq11}) can be written  as
\begin{equation}
\frac{\partial}{\partial t}\tilde{W}(s, \sigma, \Sigma, t)=\int_0^t Q(t-t')
\left[\frac{\partial}{\partial s}(2\gamma s-1)
   +\gamma\frac{\partial}{\partial\sigma}(\sigma+st)\right]\tilde{W}(s, \sigma, \Sigma, t')\,dt' \,.
\end{equation}

The difference in the two approaches of Barkai and Silbey and Friedrich et al.
can be traced back to the Langevin process (\ref{internal}). Integration
yields for instance
\begin{equation}
\sigma(t)= \int_0^t dt' s(t')= \int_0^{s(t)} ds' s' \eta(s') \qquad ,
\end{equation}
where $\eta(s)$ is a stochastic process. Since one has to integrate
over the product $s \eta(s)$ a 
stochastic interpretation of this integral is needed. 
A similar situation arises 
in stochastic processes involving multiplicative white
noise sources, where different interpretations of similar
integrals have been
given by Ito and Stratonovich (for a discussion see e.g. \cite
{Risken}). We conjecture that
different interpretations of this integral and the integral arising for
the variable $\Sigma(t)$ lies at the origin of the two different fractional
equations for the probability distributions $W(s,\sigma,\Sigma,t)$.

{\em Conclusions.} -- 
Three different types of fractional generalizations of the KFP equation
describing anomalous diffusion of inertial particles can be found in the
literature. Based on the idea of subordination, which is equivalent
to the introduction of an intrinsic, fluctuating time, we have clarified
the meaning of these different equations. Whereas in the approach of
Metzler and Klafter \cite{Metzler3} - \cite{Metzler4} both 
position and velocity depend on the intrinsic time, the approach of
Barkai and Silbey \cite{BaSil} and Friedrich et. al. \cite{Eule}, 
\cite{CTRW} assumes that only the velocity is subjected
to the subordination procedure. We conjecture that
the difference between the approaches of Barkai and Silbey and 
Friedrich et al. is due to different interpretations of stochastic 
integrals.

\end{document}